\documentclass[10pt,twocolumn,letterpaper]{article} 

\usepackage{ol2}
\usepackage{hyperref}
\hypersetup{colorlinks=true,citecolor=blue,linktoc=page}
\usepackage{amsmath,amssymb,graphicx}
\usepackage[compact]{titlesec}
\usepackage[small]{caption}
\usepackage{textcomp}

\begin{document}
\twocolumn[
\title{Multiobjective optimization in integrated photonics design}
\author{Denis Gagnon, Joey Dumont and Louis J. Dub\'e}
\affiliation{D\'epartement de physique, de g\'enie physique et d'optique \\Facult\'e des Sciences et de G\'enie, Universit\'e Laval, Qu\'ebec G1V 0A6, Canada}
\email{Corresponding author: ljd@phy.ulaval.ca}
\begin{abstract}
We propose the use of the parallel tabu search algorithm (PTS) to solve combinatorial inverse design problems in integrated photonics. To assess the potential of this algorithm, we consider the problem of beam shaping using a two-dimensional arrangement of dielectric scatterers. The performance of PTS is compared to one of the most widely used optimization algorithms in photonics design, the genetic algorithm (GA). We find that PTS can produce comparable or better solutions than the GA, while requiring less computation time and fewer adjustable parameters. For the coherent beam shaping problem as a case study, we demonstrate how PTS can tackle multiobjective optimization problems and represent a robust and efficient alternative to GA.
\end{abstract}

]

Silicon integrated optical chips offer enormous potential for practical applications. The capability to design and manufacture various planar integrated photonics components such as waveguides \cite{Frandsen2004}, beam-splitters \cite{Pottier2006} and slow-light devices \cite{Baba2008} has increased considerably in recent years. 
This broad spectrum of functionalities is enabled by the interplay of in-plane reflection and interference processes caused by the presence of scattering elements such as an arrangement of holes in a two-dimensional pattern. The arrangements can range from periodic -- for instance in 2D photonic crystals -- to aperiodic \cite{DalNegro2012, Vardeny2013} or even completely disordered \cite{Wiersma2013}. A frequently arising design issue in integrated photonics is to determine the scatterers' configuration required to achieved a given functionality. This class of NP-hard inverse design problems is often approached using \textit{metaheuristics}, optimization algorithms based on empirical rules for exploring large solution spaces \cite{Talbi2009}. 

The genetic algorithm (GA), a nature-inspired evolutionary method, is perhaps the most widely used metaheuristic in the field of optics and photonics \cite{Weile1997, Boxwell2004, Sanchis2004, Vukovic2010, Sanchez-Serrano2012, Marques-Hueso2013}. Some defining features of the canonical GA are that it uses stochastic transition rules, not deterministic ones, and has no memory of past solutions \cite{Glover1995}. The escape from local minima is then achieved using the application of a random mutation operator. For large-scale optimization problems in integrated photonics (for instance a large number of parameters, or the simultaneous optimization of multiple objective functions), this results in many instances in slow convergence. Common approaches for speeding up convergence include breaking the solution space in several pieces \cite{Vukovic2010}, or using a combination of GA and local search algorithms \cite{Glover1995}.

The aim of this letter is to show that purely deterministic metaheuristics can very well be applied to large-scale photonics design problems. The optimization algorithm chosen is the parallel tabu search (PTS), a deterministic algorithm which involves fewer adjustable parameters than the GA. The performance of PTS is compared to the standard GA for a case study, namely the inverse problem of beam shaping using a two-dimensional arrangement of dielectric scatterers \cite{Gagnon2012}. As a further illustration, we show that this algorithm is also well suited to inverse problems involving the simultaneous optimization of more than one attribute. More specifically, we apply PTS to the \emph{coherent beam shaping} problem, in other words the generation of a beam of controlled phase and amplitude profile.

\textbf{Beam shaping problem definition}. We will address a model inverse problem, namely beam shaping using a photonic crystal lattice. Consider a finite-size arrangement of air holes in a high-index dielectric core. The problem consists in finding a lattice configuration which, when illuminated with an arbitrary input beam, produces a scattered field that matches a desired profile in a given plane. In two dimensions, the beam shaping problem can be formulated as the minimization of the following objective function \cite{Dickey2005}
\begin{equation}\label{eq:g1}
g_1 = \dfrac{\int \big||u(x_0,y)|^2 - |\bar{u}(x_0,y)|^2 \big| dy}{\int |\bar{u}(x_0,y)|^2 dy } 
\end{equation}
where $x_0$ is the location of the target plane, $u(x_0,y)$ is the computed EM field on the target plane, $\bar{u}(x_0,y)$ is the desired beam at the device output (the $x-$axis is the beam propagation axis). The parameters to optimize can be defined as the geometry of the scatterers' arrangement. A given combination of scatterers is termed a \emph{solution}, or a \emph{configuration}. For a given configuration, the resulting beam $u(x_0,y)$ can be computed using a generalized Lorenz-Mie theory \cite{Nojima2005}. For definiteness, we set a basic lattice geometry and only allow the scatterers to be present or absent. Consequently, the optimization problem is a combinatorial one. This means individual solutions can be encoded via vectors of bits, the length of each vector being equal to the number of available scattering sites \cite{Sanchis2004, Vukovic2010}. 

\begin{figure}
\centering
\includegraphics{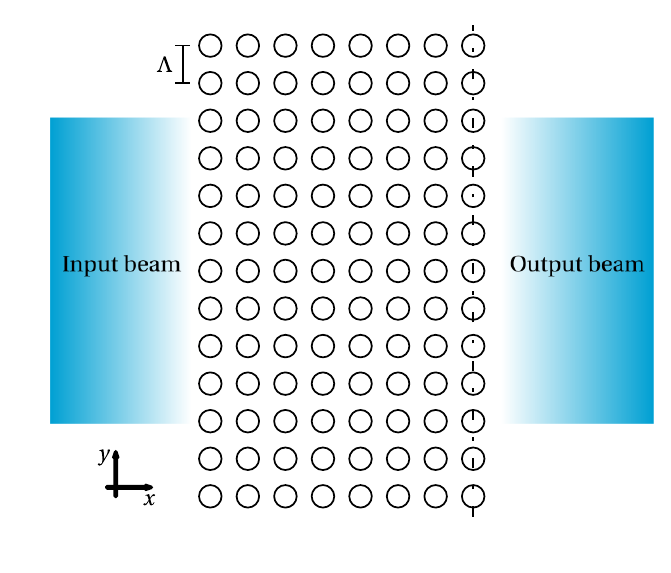}
\caption{(Color online) Basic photonic lattice configuration for the beam shaping problem. The dotted line indicates the plane used for the computation of the desired beam profile.}\label{fig:geometry}
\end{figure}

This beam shaping problem was recently tackled using a standard implementation of the GA \cite{Gagnon2012}. In this previous work, the basic geometry is a $13 \times 8$ 
square grid of scatterers with mirror symmetry, resulting in $2^{56}$ 
possible configurations (see fig. \ref{fig:geometry}). While the GA is successful in finding very acceptable solutions to this inverse problem, the minimization of $g_1$ does not take into account the phase profile of the beam, only the amplitude, or irradiance distribution. As a result, the optimized beams sometimes exhibit large transverse phase fluctuations, which in turn result in a poor field depth. This is a major impediment to applications such as atom guiding \cite{Molina-Terriza2007} and microscopy \cite{Olivier2012}, where beams with large field depths (low divergence) are needed. In order to achieve \textit{coherent beam shaping}, we must define another objective function related to phase fluctuations of the transverse profile. Consider the following integral
\begin{equation}
g_2 = \dfrac{\int \big| \mathrm{Im} [ u(x_0,y) e^{-i\phi(x_0,0)} ]  \big|^2 dy}{\int |\bar{u}(x_0,y)|^2 dy }
\end{equation}
where $\tan \phi(x,y) = \mathrm{Im}  [ u(x,y) ] / \mathrm{Re}  [ u(x,y) ]$. The value of $g_2$ is zero for a collimated beam (plane phase front), and increases with the number of oscillations in the phase front. The set of attributes $g_1$ and $g_2$ constitutes a \textit{multiobjective optimization problem} (MOP), which must be solved by sampling the set of optimal solutions, commonly known as the Pareto set \cite{Talbi2009}.

The main computational difficulty associated with MOPs lies in the fact that all Pareto solutions of a $p-$objective problem are necessarily solutions of the same problem with a larger number of objectives \cite{Talbi2009}. Consequently, the number of objective function evaluations needed to solve a MOP is significantly greater than in the single-objective case. Because of that higher computation cost, it is beneficial to use a metaheuristic algorithm that yields better solutions than the GA for a given number of objective function evaluations. However, one would want to preserve the innate ability of the GA to sample broad areas of the solution space. In light of this observation, we propose the use of the parallel tabu search (PTS) consisting in an ensemble of individual tabu search processes exploring the solution space in a parallel fashion \cite{Crainic1997}. 

\textbf{Performance assessment of PTS.} The tabu search is a deterministic local search algorithm first proposed by Glover in the late 1980s \cite{Talbi2009, Glover1989}. One iteration of a tabu search process begins with the evaluation of the objective function in the \emph{neighborhood} of the current solution. The algorithm then proceeds to the best possible neighbor (best possible value of the objective function) that is not prohibited by the \textit{tabu list}. This list of forbidden moves constitutes the short-term memory of the algorithm and prevents a cyclic search in the solution space. Its length $L$ may be kept constant or dynamically adjusted as the algorithm progresses. In our parallel implementation of the tabu method, we begin by generating a diverse ``population'' of solutions using a method known as simple sequential inhibition \cite{Talbi2009}. An individual tabu search process then begins working on a member of the initial ``population'' until a stopping criterion is met (typically 
a fixed number of iterations). 
Since each process acts in a local and deterministic way, the goal of the parallel implementation is to provide a broad sampling of the solution space, as does the GA.

To compare the performance of PTS versus the GA, we apply both algorithms to the incoherent beam shaping problem mentioned earlier. We only optimize for $g_1$ (see eq. \ref{eq:g1}) using the basic scatterer geometry shown in fig. \ref{fig:geometry}. The diameter of all air holes holes is set to $D = 0.6 \Lambda$, where $\Lambda$ is the lattice constant. We use an effective index $n = 2.76$, corresponding to a thin silicon slab at $\lambda \sim 1.5$ \textmu m \cite{Chutinan2000}. Although the input and output beams may be arbitrary, we prescribe our incident beam as a TM-polarized non-paraxial Gaussian beam with a half-width $w_0=2.5\Lambda$ and a wavenumber $k_0 = 1.76 / \Lambda$. Moreover, a mirror symmetry across the $x$ axis is taken into account, resulting in $2^{56}$ possible solutions, or $\sim 7 \times 10^{16}$. The generalized Lorenz-Mie method used to compute the scattered field $u(x,y)$ is detailed in refs. \cite{Nojima2005, Gagnon2012}.

\begin{figure}
\centering
\includegraphics{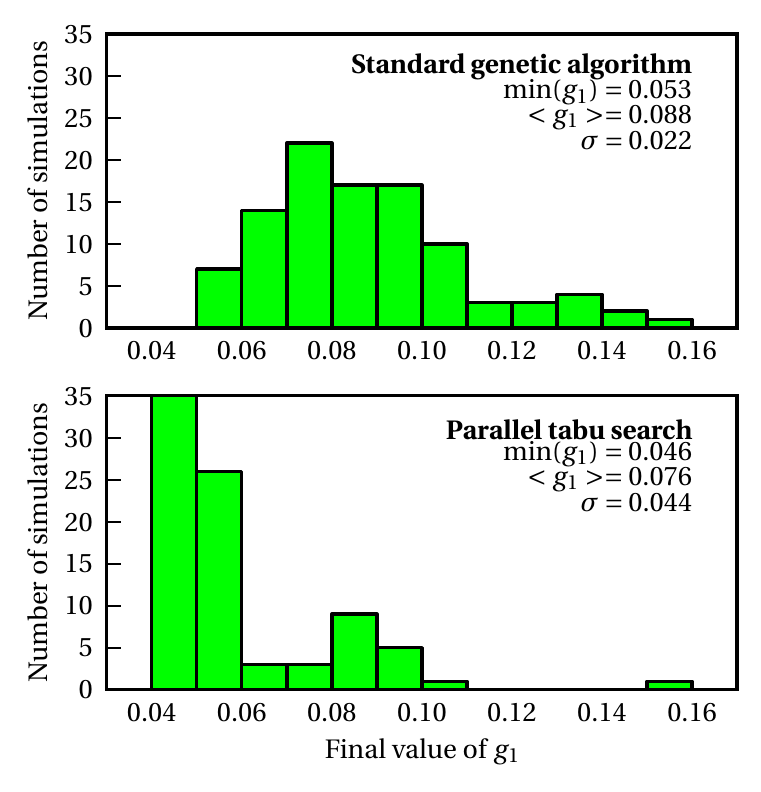}
\caption{(Color online) Comparison of the GA and PTS algorithms applied to the incoherent beam shaping problem. Each simulation represents 5000 generations/iterations, with a similar computational cost. 100 simulations are shown for each algorithm.}\label{fig:hist}
\end{figure}

The parameters of the canonical GA are set following the guidelines of Vukovic \emph{et al.} \cite{Vukovic2010}. More specifically, we use roulette wheel sampling, random mutations with probability $p_m = 0.002$, uniform crossover with probability $p_c = 0.2$, and elitism. The generation size is set to 200 individuals. On the other hand, the only PTS parameter to be specified by the user is the tabu list length $L$. In this work, we use a fixed length of $L = 2.5 \sqrt{N_n}$, where $N_n$ is the number of neighbors of a given solution. In this case, $N_n$ is also equal to the number of available scattering sites. We launched 100 GA processes and 100 parallel tabu search processes, each for 5000 iterations (or generations). Since PTS is deterministic, each iteration implies no more than $N_n = 56$ objective function evaluations, whereas we found that each generation of the GA implied an average of 60 objective function evaluations (we keep the values for the best solution in memory). This means that the 
run-time of each algorithm is similar given our choice of parameters. The minimal values of $g_1$ for each optimization algorithm are presented in fig. \ref{fig:hist}. The results show that the solutions found by PTS are more optimal on average for an equivalent computation time. Moreover, some solutions found by PTS were 
inaccessible to the GA. Therefore, for combinatorial optimization in integrated photonics, PTS may be a better choice. This is similar to the performance gain of tabu search when applied to timetable scheduling problems \cite{Chu1999}. PTS is also appealing because it involves very few adjustable parameters and its implementation is more straightforward than that of the GA.  

Furthermore, and remarkably, the configurations computed via PTS exhibit a power conversion efficiency ranging from 70 \% to 80 \%, similar to those obtained with the GA, and merely 10--20 \% lower than arrangements \emph{specifically} designed for high efficiency. Additional discussion and comparison to other integrated devices can be found in \cite{Gagnon2012}.

\begin{figure*}
\centering
\includegraphics{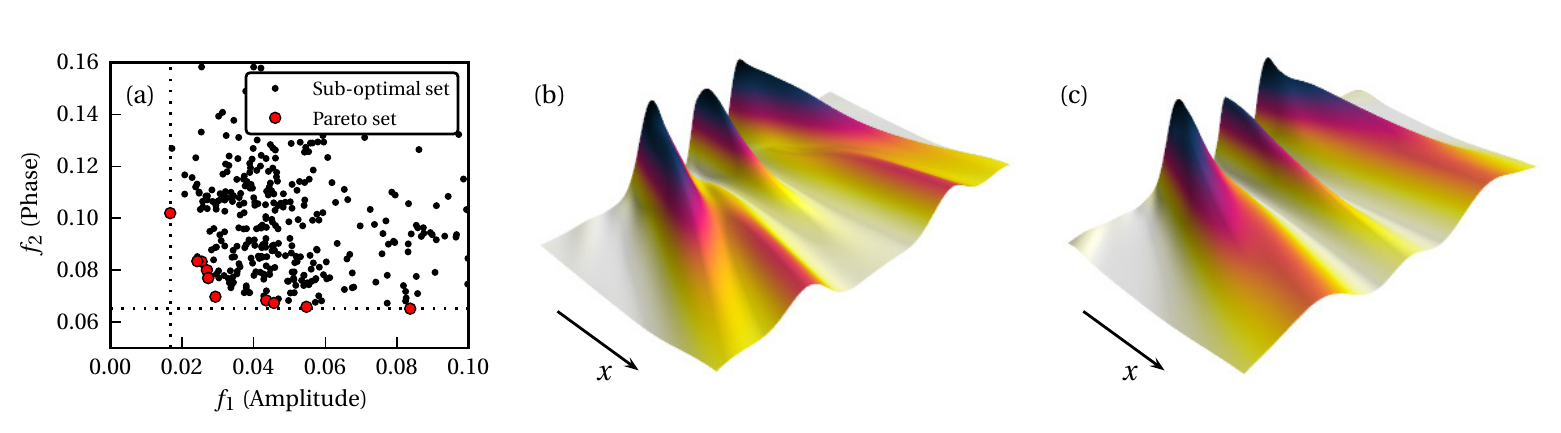}
\caption{(Color online) Multiobjective optimization results. (a) Sampling of the Pareto front for the coherent beam shaping problem. The dotted lines indicate the best possible value for each of the two objectives separately. (b) Hermite-Gauss beam profile with the lowest possible value of $f_1$ ($f_1=0.0167,f_2=0.1019$). (c) Best possible trade-off between the two objectives ($f_1=0.0255,f_2=0.0833$). Since the phase is controlled, the Hermite-Gauss profile shape is preserved over a greater propagation distance. This can be seen in the number of ridges in the transverse profile (arbitrary intensity units). The lattice configurations producing these beams are made of $N_s = 56$ (b) and $N_s = 52$ (c) scattering sites whose explicit positions can be found in \cite{Gagnon}.}\label{fig:multi}
\end{figure*}

\textbf{Multiobjective results and discussion.} Having assessed the potential of the PTS optimization algorithm, we now center our attention on the solution of the MOP described by the two objective functions $g_1$ and $g_2$. The solution to a MOP is not a single solution, but rather a set of solutions called \textit{Pareto optimal}. A solution is Pareto optimal if it is not possible to improve a given objective without deteriorating at least another \cite{Ehrgott2005}. MOPs arise in various areas of engineering and science, such as microprocessor design \cite{Sengupta2011}, medical physics \cite{Holdsworth2011}, chaotic systems \cite{Tang2011} and accelerator physics \cite{Hofler2013}. The simplest way to solve a MOP is the weighted-sum method (or aggregation method). Basically, one recasts a $p-$objective problem into a single-objective one in the following way \cite{Ehrgott2005}
\begin{equation}
\min_{\xi \in \mathcal{\varXi}} \sum_{i=1}^{p} \alpha_i f_i(\xi)
\end{equation}
where $\xi$ is a solution, $\varXi$ is the solution space, and $f_i = g_i / g_i^{\mathrm{max}}$. Objective functions must be normalized with respect to a heuristic upper bound $g_i^{\mathrm{max}}$ to ensure that all objectives are commensurate.  The Pareto front (location of the set of optimal solutions) is then sampled by solving several different single-objective problems using different values of the weights $\alpha_i$. This has the effect of increasing or decreasing the relative importance of each different objectives, thereby steering the search towards different regions of the Pareto front \cite{Talbi2009, Ehrgott2005}. In our case, this implies running several PTS processes using different values of the relative weights $\alpha_i$.

Using the weighted-sum method, we perform the simultaneous optimization of the amplitude and the phase of a order 2 Hermite-Gauss beam. The geometry used is the same as described above, except that the square lattice is somewhat larger, $13 \times 10$ scatterers, for a total of 130 possible scattering sites. Accounting for symmetry, this results in $2^{70}$ possible solutions, or $\sim 10^{21}$. We set the values $g_1^{\mathrm{max}} = 1$, $g_2^{\mathrm{max}} = 10$ and the restriction $\alpha_1 + \alpha_2 = 1$. The sampling of the Pareto front is performed using 7 different values of $\alpha_2 \in [0.0,0.425]$. For each of those values, 48 tabu search processes are performed in parallel. This set of search processes yields a number of final solutions, out of which we extract the Pareto optimal set (i.e. those solutions for which there is no solution found that is characterized by 
a lower value of both $g_1$ and $g_2$). The resulting Pareto front is shown in fig. \ref{fig:multi}a.

Once the Pareto front is sampled, the ``optimality'' of the solutions is to be evaluated \textit{a posteriori} depending on the preferred application.  In other words, it is up to the end-user, or decision maker, to determine what is the best trade-off between the predefined objectives. In our case, we are interested in generating beams with a large field depth. As illustration, the configuration in fig. \ref{fig:multi}b offers the most accurate reproduction of a Hermite-Gauss beam profile (smallest obtained value of $f_1$). However, the non-uniformity of the phase front results in a poor field depth. On the other hand, the configuration in fig. \ref{fig:multi}c exhibits a better field depth, keeping a Hermite-Gaussian profile over a greater distance. This solution would likely have been ``missed'' in the single-objective case. This last point is crucial in the optimization problem. In selecting a multi-objective versus a single-objective calculation, one must keep in mind that the former offers a much 
greater diversity and density of solutions. For instance, we found that replacing the integrand of the single-objective function $g_1$ by  $|u(x_0,y) -\bar{u}(x_0,y)|^2$ for the coherent problem, we could only reach a much smaller and less optimal subset of solutions.

\textbf{Conclusion.} In summary, we propose the use of the PTS algorithm for combinatorial optimization problems in integrated photonics. We show that PTS finds some optimal solutions faster than the standard GA for the specific problem of beam shaping using a 2D photonic lattice. Moreover, the tabu method involves fewer adjustable parameters, allowing for a straightforward implementation. Using this improved algorithm, we have reported the possibility to control the coherent profile (amplitude and phase) of the output beam. Our results show that multiobjective optimization in integrated photonics design is within reach and that a PTS algorithm offers an efficient alternative to the standard GA.

We are grateful to Alain Hertz for introducing us to the world of metaheuristics. The authors acknowledge financial support from the Natural Sciences and Engineering Research Council of Canada (NSERC) and computational resources from Calcul Qu\'ebec.

\bibliographystyle{ol}

\end{document}